\documentstyle[aps,twocolumn, floats]{revtex}
\newcommand{\be}{\begin{equation}}
\newcommand{\ee}{\end{equation}}
\input psfig.sty
\begin{document}
\draft
\title{ Single-hole dynamics in the half-filled two-dimensional 
        Kondo-Hubbard model.}
\author{M. Feldbacher$^\dagger$, C. Jurecka$^\star$, F. F. Assaad$^{\dagger,+}$ and W. Brenig$^\star$}

\address{$^\dagger$ Institut f\"ur Theoretische Physik III, 
   Universit\"at Stuttgart, Pfaffenwaldring 57, D-70550 Stuttgart, Germany. }

\address{$^\star$ Institut f\"ur Theoretische Physik,
Technische Universit\"at Braunschweig, D-38106 Braunschweig, Germany.}

\address{$^+$  Max Plank institute for solid state research, Heisenbergstr. 1,
D-70569, Stuttgart}

\date{\today}
\maketitle

\begin{abstract}
We consider the Kondo lattice model in two dimensions at half filling. In
addition to the fermionic hopping integral $t$ and the superexchange coupling
$J$ the role of a Coulomb repulsion $U$ in the conduction band
is investigated. We find the model to display a magnetic order-disorder
transition in the $U$-$J$ plane with a critical value of $J_c$ which is 
decreasing as a function of $U$.
The single particle  spectral function $A(\vec{k},\omega)$ is computed across 
this transition. For all values of $J > 0$, and apart from shadow features 
present in the ordered
state, $A(\vec{k},\omega)$ remains insensitive to the magnetic phase transition
with the first low-energy hole states residing at momenta $\vec{k} = 
(\pm \pi, \pm \pi) $.  As $J \rightarrow 0$  the model maps onto the
Hubbard Hamiltonian. Only in this  limit, the  low-energy spectral weight
at $\vec{k} = (\pm \pi, \pm \pi)$ vanishes with first electron
removal-states emerging at wave vectors on the magnetic Brillouin zone
boundary. Thus, we conclude that (i) 
the local screening of impurity spins determines the low energy behavior
of the spectral function and (ii) one cannot deform continuously the spectral 
function of the Mott-Hubbard insulator at $J=0$ to that of the Kondo insulator 
at $ J > J_c$. Our results are based on both, $T=0$ Quantum Monte-Carlo
simulations and a bond-operator mean-field theory.  
\end{abstract}

\section{Introduction}\label{sec1}
Starting from the seminal work of Brinkman and Rice~\cite{Brinkman70}
 the single-hole
dynamics in correlated insulators has remained to be an intriguing issue
with many open questions yet to be clarified. In this respect Kondo
lattice systems at half filling provide for a case in which the single
particle dynamics can be studied continuously across genuinely
distinct correlation induced insulating phases, i.e., Mott-Hubbard and
magnetic insulators as well as Kondo insulators. Of particular interest
in this situation is to understand i) which properties of the correlated
insulator, i.e., long range magnetic order or local effects determine the 
functional form of the quasiparticle dispersion relation and  more 
specifically  ii) if it is possible to continuously deform 
the spectral function of the  Mott Hubbard insulator to that of the 
Kondo insulator. In order to answer these questions, it is the purpose of
this paper to consider a Kondo lattice model (UKLM) on a two-dimensional square
lattice  with an additional local Coulomb repulsion $U$ between the conduction
electrons
\begin{eqnarray}
\label{KLMU}
H   = &  & \sum_{ \vec{k}, \sigma } \varepsilon(\vec{k})
              c^{\dagger}_{\vec{k},\sigma} c_{\vec{k},\sigma}
    + J \sum_{\vec{i}}
    \vec{S}^{c}_{\vec{i}} \cdot \vec{S}^{f}_{\vec{i}} +  \nonumber \\
    & &  U \sum_{\vec{i}} \left( n_{\vec{i},\uparrow} - 1/2\right)
                     \left( n_{\vec{i},\downarrow} - 1/2\right).
\end{eqnarray}
The unit cell, $\vec{i}$, contains a localized orbital and an extended
conduction band state. In the Kondo
limit, charge fluctuations on the localized orbital are suppressed 
with only the spin degrees of freedom remaining,
$\vec{S}^{f}_{\vec{i}} = \sum_{s,s'}  f^{\dagger}_{\vec{i},s}
\vec{\sigma}_{s,s'} f_{\vec{i},s'}/2$, where $\vec{\sigma} $ are
Pauli spin-$1/2$ matrices and
$f^{\dagger}_{\vec{i},s}$ are fermionic operators which satisfy the constraint
$\sum_s f^{\dagger}_{\vec{i},s}  f^{}_{\vec{i},s}= 1 $. Conduction band
electrons of spin $z$-component $\sigma$ are
created by  $c^{\dagger}_{\vec{i},\sigma}$ where $n_{\vec{i},\sigma}=
c^{\dagger}_{\vec{i},\sigma}c^{}_{\vec{i},\sigma}$ is the conduction
band density for spin $z$-component $\sigma$. The extended orbitals overlap
to form a band with a dispersion $ \varepsilon(\vec{k}) = -2 t (\cos(k_x) +    
\cos(k_y) )$ assuming a nearest-neighbor (NN) hopping integral $t$. The Coulomb
repulsion $U$ is taken into account by the Hubbard interaction term.

At half-filling and for the particular conduction band  structure chosen
the UKLM is an insulator for all values of $U$ and $J$
~\cite{Imada_rev,Tsunetsugu97_rev,Assaad99a,Capponi00}.
Specifically, 
as  $ J \rightarrow 0$ the UKLM maps onto the Hubbard model. The latter is in
a Mott insulating phase, which however due to nesting on the square lattice
is masked by an antiferromagnetic spin density wave with no spin gap and
a charge gap $\propto U$
in the strong coupling limit. As $J/U \rightarrow \infty$ the Hubbard repulsion
can be neglected relative to the exchange scattering and the model maps onto
the pure Kondo lattice model (KLM) with $J/t \gg 1$.
In this limit, the ground state is a Kondo insulator with spin  
($ \Delta_s$)  
and charge ($ \Delta_c$) gaps satisfying $\Delta_c > \Delta_s$ 
\cite{Tsunetsugu97_rev}.
In both of the aforementioned limiting cases, the single particle spectral
function displays very different behavior. The limiting ground state for $J/t
\rightarrow\infty$, i.e. 
the Kondo insulator, is a direct product of Kondo singlets on the 
$f$-$c$ bonds of the unit cell. Adding a hole into the conduction band will
break a singlet and leads to a hole dispersion 
relation $\tilde{\varepsilon}(\vec{k})=3 J/4 + t (\cos(k_x) + \cos(k_y)) $
in first order perturbation theory in $t/J$~\cite{Tsunetsugu97_rev}. Hence the first
electron removal-states ('Fermi points') occur at wave vectors
$\vec{k} = (\pm \pi,\pm \pi)$ within the Brillouin zone. At present
the precise form the the single particle spectral function for the Mott insulating
state is still unknown. Yet, various numerical and analytical 
approaches  confirm  that the first electron removal-states are located at
$\vec{k}$-points on the boundary of the magnetic Brillouin zone, i.e., at
$\vec{k} = (0,\pm \pi), (\pm \pi, 0)$ and   
$\vec{k} = (\pm \pi/2, \pm \pi/2)$. In order to shed light onto this situation
we can fix $U$ and, as a function of $J/t$, drive the system through a
magnetic quantum phase transition at $J = J_c(U)$
from the Kondo insulator for $J/t >> 1$ into the  Mott insulating state 
for $J \rightarrow  0$. Along this path we compute the spectral function 
$A(\vec{k},\omega)$, both, {\it exactly} by
using Quantum Monte Carlo methods and approximately using a
bond-operator mean field theory. Based on our findings we will argue that the
low energy features 
of the spectral function are insensitive to the quantum phase transition. 
Speaking differently, the low energy hole-states are found at $\vec{k} = 
(\pm\pi,\pm\pi) $ for all values of $J > 0$. It is only at $J =0$ that 
the spectral
weight of 
the low energy feature at $\vec{k} = (\pm\pi,\pm\pi)$  vanishes to produce 
the single-hole 
dispersion relation of the Hubbard model.  Thus our main results are (i) that 
the local screening of the $f$-spins dominates the low energy features of the 
spectral function and (ii) that there is no continuous path from the 
 Kondo to the 
Mott-Hubbard insulator in this specific model.

The paper is organized as follows.
In the next section, we briefly outline the 
Quantum Monte-Carlo (QMC) method as 
well as the bond-operator-mean field theory.   
In  section \ref{sec3}, we first 
discuss the magnetic phase diagram of the UKLM model. We map out the 
critical line in the $U$-$J$ plane to show that $J_c$ is a monotonically
decreasing function of $U$. Comparison
of the QMC and  mean-field results prove to be very satisfactory. 
Having determined $J_c$ as a function of $U/t$ we then focus on the single  
particle  spectral function and detail the aforementioned results (i) and (ii).
Section \ref{conclude} is devoted to conclusions.

\begin{section}{Methods}
\label{sec2}
\begin{subsection}{Quantum Monte Carlo}

We have used the projector  auxiliary field Quantum-Monte-Carlo
(PQMC) method to investigate the UKLM model. This method is based 
on the projection equation
\[
\frac{\left\langle \Psi_{0}\right|  O\left|  \Psi_{0}\right\rangle
}{\left\langle \Psi_{0}\right|  \left.  \Psi_{0}\right\rangle }=\lim
_{\theta\rightarrow\infty}\frac{\left\langle \Psi_{T}\right|  e^{-\theta
H} Oe^{-\theta H}\left|  \Psi_{T}\right\rangle
}{\left\langle \Psi_{T}\right|  e^{-2\theta H}\left|  \Psi
_{T}\right\rangle }%
\]
where $\left|  \Psi_{T}\right\rangle $ is required to be non--orthogonal to
the ground state $\left|  \Psi_{0}\right\rangle$.  Details of how to implement
the PQMC method without generating a sign problem in the particle-hole
symmetric case, i.e. at half-filling, have been introduced and extensively
described for the KLM in Refs. \cite{Assaad99a,Capponi00}. We have also used a
new and efficient method 
\cite{Feldbach00} to calculate imaginary  time displaced Greens functions 
within the PQMC formalism.   From the technical point of view no complications arise when 
introducing the  Hubbard term into the KLM. Hence we refer the reader to Refs. 
\cite{Assaad99a,Capponi00,Feldbach00} for a detailed description of the method.

\end{subsection}
\begin{subsection}{Mean field theory}
For an approximate description of the UKLM we apply a mean field 
theory similar to the one proposed for the pure KLM in  \cite{Jurecka01}, 
where further details can be found. We represent the local Hilbert space 
consisting of one $f$ electron and additionally up to two itinerant electrons by
 applying the following operators onto the vacuum $|0\rangle$ of an empty site
\begin{eqnarray}
s^\dagger|0\rangle&=&\frac{1}{\sqrt{2}}(
c^\dagger_\uparrow f^\dagger_\downarrow+f^\dagger_\uparrow 
c^\dagger_\downarrow)|0\rangle \nonumber\\
t^\dagger_x|0\rangle&=&\frac{-1}{\sqrt{2}}(c^\dagger_\uparrow 
f^\dagger_\uparrow-c^\dagger_\downarrow f^\dagger_\downarrow)
|0\rangle \nonumber\\
t^\dagger_y|0\rangle&=&\frac{i}{\sqrt{2}}(c^\dagger_\uparrow 
f^\dagger_\uparrow+c^\dagger_\downarrow f^\dagger_\downarrow)
|0\rangle \nonumber\\
t^\dagger_z|0\rangle&=&\frac{1}{\sqrt{2}}(c^\dagger_\uparrow 
f^\dagger_\downarrow+c^\dagger_\downarrow f^\dagger_\uparrow)
|0\rangle \nonumber\\
a^\dagger_\sigma|0\rangle&=&f^\dagger_\sigma|0\rangle \nonumber\\
b^\dagger_\sigma|0\rangle&=&c^\dagger_\uparrow c^\dagger_\downarrow 
f^\dagger_\sigma|0\rangle.
\label{e1}
\end{eqnarray}
The $s$ and $t$ operators are equivalent to the so-called bond operators
of\cite{Sachdev90} and are assumed to obey {\em bosonic} commutation relations.
The {\em fermionic} operators $a$ and $b$ have been introduced first
in\cite{Eder97,Eder98} and label states with one or three electrons per site. 
In order
to suppress unphysical states a constraint of no double occupancy
\be
s^\dagger_j s^{\phantom\dagger}_j+\sum_\alpha t^\dagger_{\alpha,j} 
t^{\phantom\dagger}_{\alpha,j}+\sum_\sigma a^\dagger_{\sigma,j}
a^{\phantom\dagger}_{\sigma,j}+\sum_\sigma b^\dagger_{\sigma,j}
b^{\phantom\dagger}_{\sigma,j}=1
\label{e2}
\ee
has to be fulfilled.

Rewriting the UKLM in terms of (\ref{e1}) leads to a strongly correlated
boson-fermion model which cannot be solved exactly. To proceed we use a
mean-field approach which incorporates both, an antiferromagnetic state 
and a spin-singlet regime. The latter regime can be expressed by a condensate 
of the singlets \cite{Sachdev90}
\be
\langle s^{\phantom\dagger}_j \rangle=\langle s^\dagger_j \rangle=s.
\label{e3}
\ee
In the antiferromagnetic phase the dimers condense into a linear combination
of the singlet and one of the triplets \cite{Normand97} implying that
\begin{eqnarray}
&&\langle t^{\phantom\dagger}_{z, j} \rangle=\langle t^\dagger_{z,j} 
\rangle=m_j=(-1)^j m
\label{e4}\\
&&\langle s^{\phantom\dagger}_j \rangle=\langle s^\dagger_j \rangle=s.
\end{eqnarray}
Here $(-1)^j$ is a shorthand for '$+1(-1)$' on the magnetic A(B) sublattice.

Inserting (\ref{e1}) into the UKLM and using the mean field approximation
(\ref{e3},\ref{e4}) we obtain
\begin{eqnarray}
H&=&-\frac{t}{2}\sum_{\{i,j\},\sigma}(-sp_\sigma+m_i)(-sp_\sigma+m_{j})\times
\nonumber\\ 
&&\qquad\times(a^{\phantom\dagger}_{\sigma,i} a^\dagger_{\sigma,j}+
b^\dagger_{\sigma,i}b^{\phantom\dagger}_{\sigma,j})\nonumber+h.c.\\
&&-\frac{t}{2}\sum_{\{i,j\},\sigma}(-sp_\sigma+m_i)(sp_\sigma+m_{j})
\times\nonumber\\&&\qquad\times(-p_\sigma a^{\phantom\dagger}_{\sigma,i} 
b^{\phantom\dagger}_{-\sigma,j}+p_\sigma b^\dagger_{\sigma,i}
a^{\dagger}_{-\sigma,j})+h.c.\nonumber\\&&
-\frac{3}{4}JNs^2+\frac{1}{4}JNm_i^2\nonumber\\&&
+\sum_{i, \sigma} \mu_i (s^2+m_i^2+a^\dagger_{\sigma,i}
a^{\phantom\dagger}_{i,\sigma}+b^\dagger_{\sigma,i}
b^{\phantom\dagger}_{\sigma,i}-1)\nonumber\\&&
+\lambda \sum_{i, \sigma} (b^\dagger_{\sigma,i} b^{\phantom\dagger}_{\sigma,i}-
a^\dagger_{\sigma,i}a^{\phantom\dagger}_{\sigma,i})\nonumber\\&&
+\frac{UN}{4}-\frac{UN}{2}(s^2+m_i^2)
\label{e5}
\end{eqnarray}
where $p_{\uparrow(\downarrow)} = 1(-1)$ and we have introduced a chemical
potential $\lambda$ to set the global particle density and a local Lagrange
multiplier $\mu_i$ in order to to enforce the constraint (\ref{e2}). In the
remainder of this work we assume $\mu_i$ to be site independent, i.e.
$\mu_i=\mu$. The difference between (\ref{e5}) and the mean-field Hamiltonian
for the pure KLM \cite{Jurecka01} resides in the last line of (\ref{e5}) which
accounts for a suppression of doubly occupied conduction electron orbitals.

Diagonalizing (\ref{e5}) leads to $4$ bands $\omega_{1,2}({\vec
k})=\lambda\pm E_1({\vec k})$ and $\omega_{3,4}({\vec k})=\lambda\pm E_2({\vec
k})$ which are twofold degenerate by spin-z quantum number
\begin{eqnarray}
E_{\stackrel{\scriptstyle{1}}{2}
\stackrel{\scriptstyle{,{\vec k}}}{\phantom{}}}
&=&\sqrt{\mu^2+\frac{1}{2}\epsilon_{\vec k}^2(m^2+s^2)^2 \mp
2W_{\vec k}} \nonumber \\
W_{\vec k}&=&\sqrt{\frac{1}{4}\mu^2(m^2-s^2)^2\epsilon_{\vec k}^2+
\frac{1}{16}\epsilon_{\vec k}^4(m^2+s^2)^4}
\label{e6}
\end{eqnarray}
Here $\epsilon_{\vec k}=-2t\sum_{d=1}^D \cos k_d$. Note, that the dispersions
in (\ref{e6}) do depend on $U$, as the order parameters $s$, $m$ and $\mu$ are
functions of $U$. At half filling the lower(upper) two
bands, i.e $\omega_{2,4}$ ($\omega_{1,3}$), are completely filled(empty).

Evaluating the ground state energy and using the stationarity conditions
$\partial E/\partial s=0$, $\partial E/\partial m=0$, and $\partial E/
\partial \mu=0$ the mean-field equations in the magnetic phase ($m\neq0$)
read
\begin{eqnarray}
0&=&2J+\frac{1}{2N}\sum_{\vec k} \frac{\epsilon_{\vec k}^2 \mu^2(s^2-m^2)}
{W_{\vec k}}E^-_{\vec k}
\nonumber\\
0&=&s^2+m^2+1
-\frac{1}{2N}\sum_{\vec k} \left[ \mu E^+_{\vec k}
+ \frac{\epsilon_{\vec k}^2\mu(s^2-m^2)^2}
{4W_{\vec k}}E^-_{\vec k} \right]
\nonumber\\
0&=&-J+4\mu-2U
\nonumber\\[3pt]
&&-\frac{1}{2N}\sum_{\vec k} \left[2\epsilon_{\vec k}^2(m^2+s^2)
E^+_{\vec k} + \frac{\epsilon_{\vec k}^4(m^2+s^2)^3}
{2W_{\vec k}}E^-_{\vec k} \right]
\label{e8}
\end{eqnarray}
where $E^\pm_{\vec k}=
2(E^{-1}_{2, {\vec k}} \pm E^{-1}_{1, {\vec k}})$.
For the disordered Kondo-singlet phase ($m=0$) we get
\begin{eqnarray}
0&=&-\frac{3}{2}J+2\mu-U-\frac{1}{N}\sum_{\vec k} \frac{2\epsilon_{\vec k}^2s^2}
{\sqrt{4\mu^2+\epsilon_{\vec k}^2s^4}}\nonumber\\
0&=&s^2+1-\frac{1}{N}\sum_{\vec k}\frac{4\mu}{\sqrt{4\mu^2+
\epsilon_{\vec k}^2s^4}}.
\label{e9}
\end{eqnarray}

Fig. \ref{fig2} shows numerical solutions of (\ref{e8}) and (\ref{e9}) as 
a function of $J$ and $U$. In both cases we find a second order phase transition
between the antiferromagnetically ordered and the Kondo phase. Fig. 1 also shows
the staggered magnetizations \cite{Jurecka01} $M_{c(f)}$ of the $c(f)$ electron
\begin{eqnarray}
M_c&=&\frac{2}{N}\sum_n(-1)^n\langle S^c_{z,n}\rangle=2ms\nonumber\\
M_f&=&\frac{2}{N}\sum_n(-1)^n\langle S^f_{z,n}\rangle=\nonumber\\&=&
2ms+\frac{1}{N}\sum_{\vec k}\frac{2 \epsilon_{\vec k}^2\mu m s(s^2+m^2)}
{E_{1,{\vec k}}E_{2,{\vec k}}(E_{1,{\vec k}}+E_{2,{\vec k}})}.
\label{e9a}
\end{eqnarray}

\begin{figure}
\centerline{\psfig{file=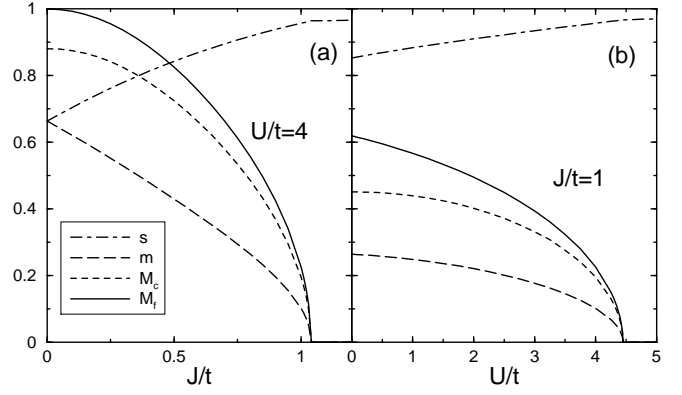,width=\linewidth,angle=-90}}
\caption{Mean-field order parameter $s$, $m$ and the staggered magnetizations
$M_c$ and $M_f$ as function of (a) $J/t$ for $U/t=4$ and (b) $U/t$ for $J=t$.}
\label{fig2}
\end{figure}

Using (\ref{e1}) we may express the spectral function $A_c({\vec k},\omega)$ of
the conduction electron $c_{\vec k}$ via a multi-particle correlation function
of the $s$, $t$, $a$ and $b$ operators. On the mean-field level however, this
simplifies into a linear combination of one-particle propagators of the $a$ and
$b$ fermions only, involving both, diagonal as well as off-diagonal contributions.
After some algebra we get
{\small
\begin{eqnarray}
&&A_c({\vec k},z)=\label{e10}\\
&&-\frac{1}{\pi}{\rm Im}\frac{(m^2+s^2)(z^2-
\mu^2)z+\epsilon_{\vec k}(z^2(s^2+m^2)^2-4m^2s^2\mu^2)}
{\mu^4+4\epsilon_{\vec k}^2m^2\mu^2s^2-z^2(2\mu^2+
\epsilon_{\vec k}^2(s^2+m^2)^2)+z^4}. \nonumber
\end{eqnarray}}
where $z=\omega+i\delta$.
\end{subsection}
\end{section}

\begin{section}{Results}
\label{sec3}

\begin{subsection}{Magnetic phase diagram}

\begin{figure}
\centerline{\psfig{file=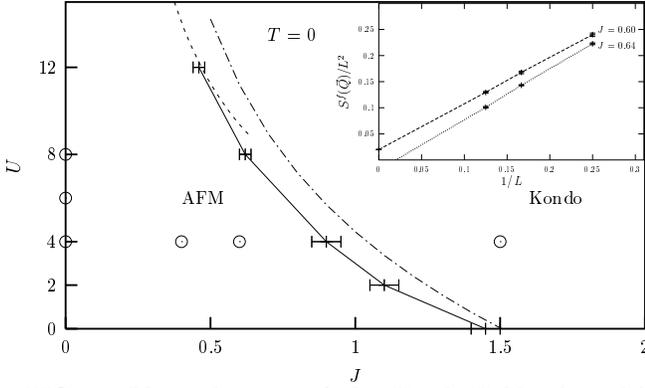,width=\linewidth,angle=0}}
\caption{Phase diagram of the Kondo-Hubbard model. Solid line: 
QMC, dot-dashed line: MF, dashed line: 
Spin Hamiltonian $H_{\rm spin}$ Eq. (\ref{Hspin}).
Circles show parameter values, where the spectral function has been evaluated.}
\label{fig3}
\end{figure}

We start the discussion of our results with the magnetic phase diagram. 
At $U=0$ the UKLM maps  onto the KLM.  In the latter, the competition 
between the  Ruderman-Kittel-Kasuya-Yosida (RKKY)  interaction and
the Kondo screening leads to a quantum phase transition
between an ordered magnetic state and the disordered singlet phase at
$J_c/t \sim 1.5 $ \cite{Doniach77,Assaad99a,Capponi00}. For 
$U/t \rightarrow \infty $ double occupancy of the conduction electron
sites is suppressed and the
model maps onto a pure spin Hamiltonian of the form:
\begin{equation}
  H_{\rm spin} = 
   J_{\parallel} \sum_{\langle \vec{i},\vec{j} \rangle } \vec{S}^{c}_{\vec{i}} 
    \vec{S}^{c}_{\vec{j}}  + J \sum_{\vec{i}}  \vec{S}^{c}_{\vec{i}} 
    \vec{S}^{f}_{\vec{i}}   
\label{Hspin}
\end{equation}
with $J_{\parallel} = 4 t^2/U$.  Hence in the limit   $U \rightarrow 
\infty $,  $J_c $ vanishes and  the ground state is  a  product of singlets
on the $f$-$c$ bonds. 

We have determined  $J_c$ as a function of the Hubbard repulsion $U$ both,
on the mean-field (MF) level and with the QMC method.
Within MF theory the staggered magnetization is given by Eq. (\ref{e9a}), 
while from the QMC method it is determined using the static spin-spin 
correlation
function
\begin{eqnarray}
S^{\alpha}\left(  \vec{j}\right)     &=&\left\langle \vec{S}_{\vec{j}}^{\alpha}\vec
{S}_{0}^{\alpha}\right\rangle \nonumber \\
S^{\alpha}\left(  \vec{q}\right)    & =&\sum_{\vec{j}}e^{i\vec{q}\vec{j}%
}S^{\alpha}\left(  \vec{j}\right),
\end{eqnarray}
$\alpha=c(f)$ labels conduction(f) electron spins $S^{c(f)}_{\vec{j}}$
and the total spin $S^{tot}_{\vec{j}}$ is given by $S^{tot}_{\vec{j}}=
S^{c}_{\vec{j}}+S^{f}_{\vec{j}}$. The staggered moment is extracted
from finite size extrapolation
\begin{eqnarray}
m^{\alpha}=\sqrt{\lim_{N\rightarrow\infty}S^{\alpha}\left(  \vec{Q}\right)/N  }  %
\end{eqnarray}
where $\vec{Q}  = (\pi,\pi) $ and $N$ is the number of unit cells.

Fig. \ref{fig3} depicts the phase diagram as a function of $J/t$ for finite $U/t$. 
The solid line refers to $QMC$ results, the dashed-dotted line shows the
mean-field results. As anticipated already by the preceding discussion of the
limiting points $U \rightarrow \infty$ and $U = 0$, the critical value $J$ is a
monotonically decreasing function of $U$. 
This can be understood as the Hubbard interaction tends to localize the conduction 
electrons leading to an effective reduction of the hopping amplitude. Hence, the
formation of local singlets is favored.
The above spin Hamiltonian (\ref{Hspin}) has been analyzed by Matsuhita and 
collaborators \cite{Matsushita97} who find a phase transition between a spin liquid
and antiferromagnetically ordered phase at $(J_\parallel/J)_c=0.71$. This leads
to $U_c=\frac{4t^2}{0.71J}$, the dashed line in fig. \ref{fig3}, in consistence 
with the results for the Kondo-Hubbard model in the {\it large} $U/t$ limit. 

Finally Fig. \ref{fig7} plots the staggered magnetization from a QMC 
scan at fixed $J$.
The broken-symmetry ground state satisfies $m^{tot}=m^f-m^c$. In QMC $m^{tot}$ was calculated 
independently and up to $U \leq 4$ the above relation is fulfilled within the errorbars. 
With increasing $U$ conduction electrons get more and more localized and their local moment
grows until it reaches the maximum of $\left\langle ( \vec{S}_{\vec{j}}^{c} )^2 
\right\rangle = 3/4$ in the strong coupling region $U>8$. The staggered moments in the small 
$U \leq 4$  region are well understood within a N\'eel picture of almost fully ordered 
$f$-spins where the small local moment of a conduction electron is anti-parallel to the impurity
spin. For larger values of $U$ dimerization becomes important which suppresses both $m^{c,f}$.

\begin{figure}
\centerline{\psfig{file=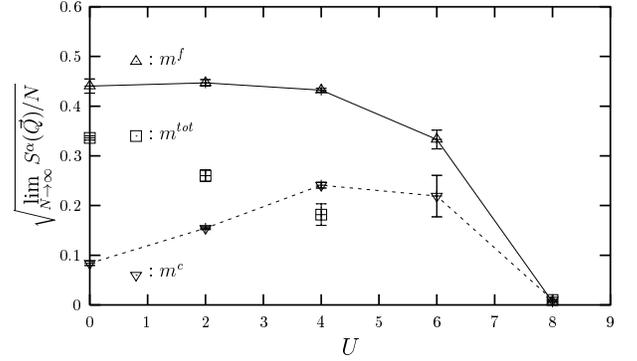 ,width=\linewidth,angle=0}}
\caption{Staggered magnetization for a fixed $J=0.6$.}
\label{fig7}
\end{figure}

\end{subsection}

\begin{figure}
\centerline{\psfig{file=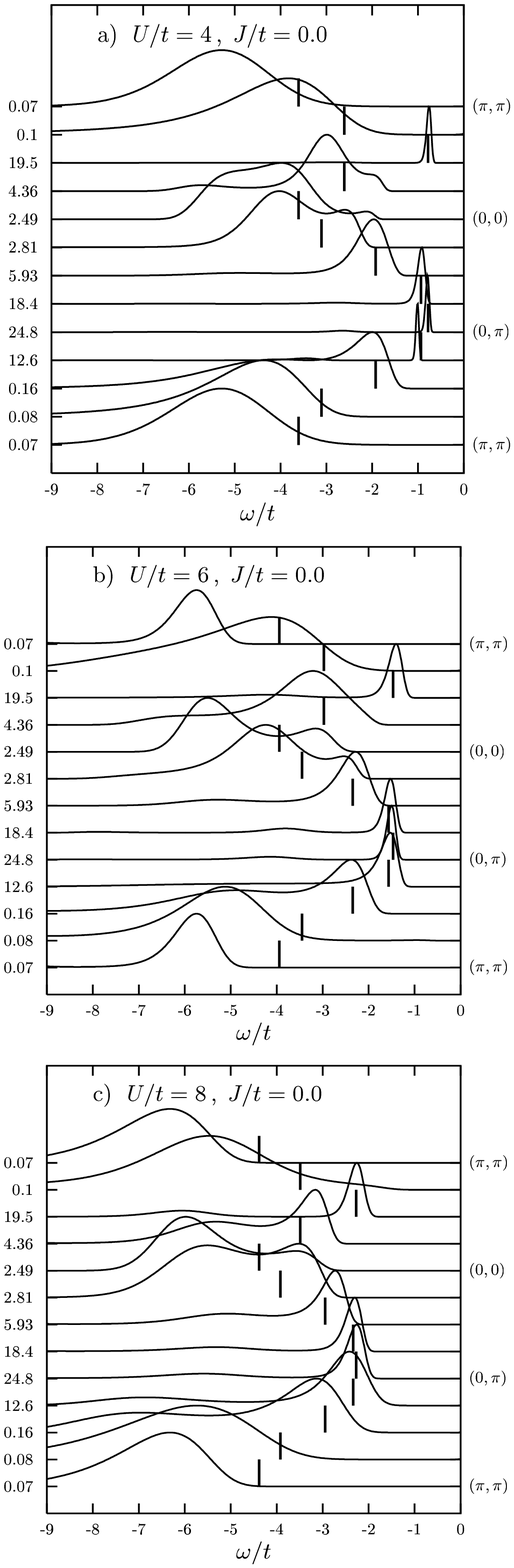,width=7cm}}
\caption{Single particle spectral function for pure Hubbard model at $U=4$, 
$U=6$ and $U=8$. For the QMC data (solid lines) we have normalized 
the maximum peak heights to unity. The numbers on the left-hand side of the 
figures correspond to the normalization factor. 
The vertical bars show the MF dispersion relation.}
\label{fig4}
\end{figure}

\begin{subsection}{Single particle spectral function.}

To study the single-hole dynamics we analyze the spectral function $A(\vec{k},
\omega)$, both using the MF expression (\ref{e10}) as well as results from the 
QMC. Within the QMC approach we first evaluate the imaginary time Greens function
\[
G_{\vec{k}}\left(  \tau\right)  =\frac{\left\langle \Psi_{0}\right|
c_{\vec{k}}^{\dagger}\left(  \tau\right)  c_{\vec{k}}\left|  \Psi_{0}\right\rangle
}{\left\langle \Psi_{0}\right|  \left.  \Psi_{0}\right\rangle }=\frac{1}{\pi
}\int_{0}^{\infty}d\omega e^{-\tau\omega}A\left(  \vec{k},-\omega\right).
\]
from which $A (  \vec{k},\omega ) $ is extracted using the maximum entropy
(ME) method \cite{Jarrell96}.

We begin with the pure Hubbard model. In Fig. \ref{fig4} we plot $A(\vec{k},
\omega)$ as obtained from QMC as well as the MF dispersion as a function of $U/t$.
While the comparison of the QMC with the MF dispersion is favorable 
one has to realize that the MF approach overestimates the quasiparticle gap.
Therefore the MF band structure in these figures results from taking only
$s$ and $m$ as obtained from the self-consistency equations (\ref{e8})
however adjusting $\mu$ such as to obtain the QMC gap at $\vec{k}=(
\frac{\pi}{2},\frac{\pi}{2})$. At weak coupling $U/t\ll 1$ we find that the
overall form of the low-energy dispersion is well reproduced by a functional
form $\pm (\Delta^2 + \epsilon(\vec{k})^{1/2}$ which is consistent with that
in a spin density wave (SDW) state. Exactly this
dispersion emerges also from the bond-operator MF theory at $J =0$
where $E_{1,\vec{k}}$ reduces to the SDW dispersion and the spectral weight
of excitations with the dispersion $E_{2,\vec{k}}$ vanishes. For $J =0$
the condensate densities for the triplet and singlet are identical, i.e.
$m = s$, which is equivalent to a N\'eel state of the $f$-spins.

At strong coupling $U/t\gg 1$ the 
Hubbard model maps approximately onto the $t$-$J_{\parallel}$ model with 
$J_{\parallel} = 4t^2/U$. Monte Carlo 
results for the latter model at $J_{\parallel}/t < 1$ show the existence of a 
quasiparticle band of width $\sim J_{\parallel}$ \cite{Brunner00b}.
This should be compared to an identical spectral feature which can be observed
in our QMC data for the Hubbard model upon enhancing $U/t$
in fig. \ref{fig4}b,c) (see also \cite{Preuss95}). 
Especially along the line from $\vec{k} = (0,\pi)$  to $\vec{k} = (0,0)$, 
this narrow  quasi particle band is clearly visible. 
In principle one should observe a similar band along
$ (0,\pi) $--$ (\pi,\pi)$, however, due to small spectral
weight in this region, we are unable to resolve this feature.                        
Of particular importance is, that for the parameters we have investigated,
the momenta of the dominant lowest energy hole-states for the Hubbard model
are found on the boundary of the magnetic Brillouin zone. For the calculations
presented in this work we have been unable  to resolve an energy difference
between the $(\pi/2, \pi/2)$ and $( 0, \pi)$ points. 

Next we turn to the UKLM at finite $J$. In fig. \ref{fig5} we show a scan of
QMC spectral functions and the MF dispersion ranging from the Kondo phase 
for $J/t = 1.5$ and
$U/t =4$ to the antiferromagnetic phase at $J/t = 0.4,0.6$ and $U/t =4$.
As for the pure Hubbard model the QMC and MF results are reasonably consistent.
From the perturbative argument for $J/t >>1$, given in section \ref{sec1}, we
expect the momenta of the dominant lowest energy hole-states to occur at
$\vec{k} = (\pm\pi,\pm\pi) $. As can be seen from fig. \ref{fig5}a) this is
in consistent, both with the QMC as well as with the MF results. Moreover
the QMC and MF dispersion agree very well.
\begin{figure}
\centerline{\psfig{file=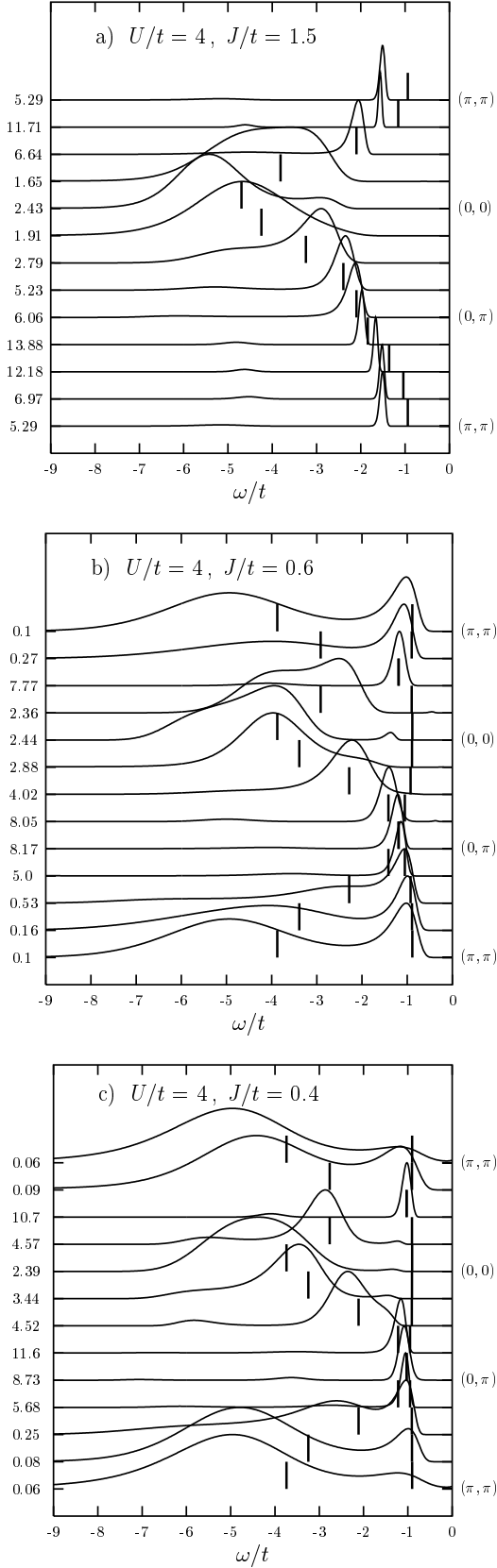,width=7cm,angle=0}}
\caption{Single particle spectra for the Kondo phase (a) and in the
antiferromagnetic phase (b,c). Vertical bars show the MF dispersion.}
\label{fig5}
\end{figure}

Lowering $J$ as in fig. \ref{fig5}a)-c) reveals the evolution of the spectral
density on going from the Kondo to the antiferromagnetically ordered phase. In
fact, as $J$ approaches zero {\em additional} bands with a dispersion similar to
the pure Hubbard case, i.e. \ref{fig4}a), develop. Yet, in the antiferromagnetically
ordered phase, but for a finite $J$ the lowest energy hole-states
are still Kondo-like, i.e. they occur at $\vec{k}=(\pm\pi,\pm\pi)$ as can be 
seen in fig. \ref{fig5}b),c). However, the weight of this excitation decreases 
continuously with decreasing $J$ and vanishes at $J=0$. 
The weight of the Hubbard-like band at $\vec{k}=(\pi,\pi)$  
increases from zero in the spin singlet phase to its maximum value at $J=0$. 
Therefore we can interpret the change in the spectral function with decreasing 
$J$ as a continuous transfer of weight from Kondo-like to Hubbard-like bands.
This shift of spectral weight renders the $J = 0$ point singular since there is 
a sudden change of the wave vector which dominates the low energy hole dynamics. 

\begin{figure}[tbh]
\centerline{\psfig{file=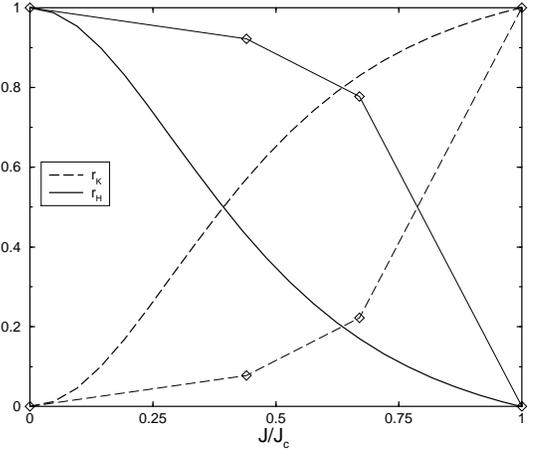 ,width=7cm,angle=-90}}
\caption{Relative spectral weight of Hubbard-band vs. Kondo-band. Thick lines:
MF. Thin lines: QMC.}
\label{fig6}
\end{figure}
\end{subsection}

These findings are corroborated by our MF results. In fig. \ref{fig6} the
relative weight $r_{H(K)}=Z_{H(K)}/(Z_H+Z_K)$ of the lower(upper)
Hubbard(Kondo)-like band at momentum $\vec{k}=(\pi,\pi)$ as obtained from
integrating $A_c(\vec{k},\omega)$
is depicted. In the QMC approach $Z_K$ results from fitting the long-time tail of
the Greens function at $\vec{k} = (\pi,\pi)$ to the form
$Z_K^{-\Delta_{qp} \tau}$ where $\Delta_{qp} $ corresponds to the quasiparticle
gap. In turn $Z_H$ is obtained
from the sum rule $ Z_K + Z_H = \pi n(\vec{k})$ assuming a two-pole structure.
Both, the QMC and the MF approximation display the same overall trend:
at $J/t=0$ the total weight is in the Hubbard-like band while with increasing $J$ 
it becomes distributed into both bands. In the Kondo phase the Hubbard band 
disappears completely. In addition fig. \ref{fig6} shows, that the MF approximation
underestimates the spectral weight in the Hubbard-like band. 

\begin{figure}
\centerline{\psfig{file=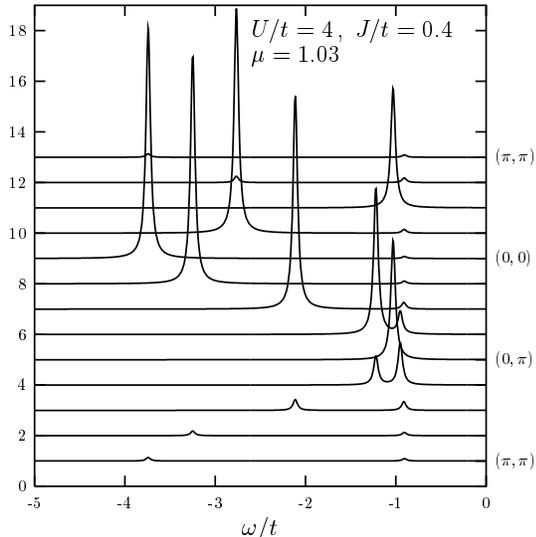,width=7cm,angle=0}}
\caption{Mean-field spectral function Eq. \ref{e10} in the antiferromagnetic phase.}
\label{fig6a}
\end{figure}

To compare the momentum dependence of the spectral weight as obtained from the
MF theory with that of the QMC fig. \ref{fig6a} depicts $A_c(\vec{k},\omega)$
from (\ref{e10}) for  $U/t=4$ and $J/t=0.4$. For visualization purpose, we have
smeared the delta-function like MF-spectrum by a finite imaginary part
$\delta=0.03$. The weight of these delta-peaks strongly varies as a function
of $\vec{k}$ having its maximum around $(0,0)$ and a very small value
in the vicinity of $(\pi,\pi)$. Again, this is consistent with the QMC data of
fig. \ref{fig5}c). Obviously, since the imaginary part of the self energy vanishes
in the MF approximation, the broadening of the  QMC spectral function is not
reproduced.  Note however, that on the QMC side pinning down the details of the
line shape is extremely challenging. 

\section{Conclusion}\label{conclude}
We have considered the single-hole dynamics in the Kondo-Hubbard model using both, 
QMC methods and a bond-operator mean-field approximation.   
Both approaches allow for similar conclusions.
The UKLM shows a magnetic order-disorder transition. At $U =0$ this transition
is triggered by the competition between the RKKY interaction and the Kondo screening
and occurs at $J_c/t \sim 1.5$. In the large $U/t$ limit the model maps onto a bilayer 
spin-model and $J_c$ scales to zero. Our results show that both limiting cases are
linked continuously and that $J_c$ is a monotonically decreasing function of $U$.
Hence as far as $J_c$ is concerned the dominant effect of the Hubbard interaction
$U$ is to localize the conduction electrons which favors screening of the localized 
spins. 

The single particle spectral function was shown to be insensitive to the 
magnetic phase transition. Irrespective of $U$ and $J > 0$  the dominant
low energy hole-states are found at the momenta $\vec{k} =(\pm\pi,\pm\pi)$. 
These excitations originate from the screening of the magnetic impurities and
hence are local. In the ordered phase pronounced shadow features can be observed.
As $J \rightarrow 0$, the spectral weight 
in the Kondo-like low  energy band in the vicinity of $\vec{k} = (\pm\pi,\pm\pi) $
vanishes and is transfered to higher energy Hubbard-like
bands. In the $(U,J)$-plane the Hubbard-line, i.e. $J = 0$, is 
singular since the localized spins decouple and lead to a macroscopically
degenerate ground state. In turn, the evolution of the spectral function is
discontinuous
in as such that at $J = 0$ there is a sudden change of the wave vector which 
dominates the low energy hole dynamics. In this sense the model shows no
continuous path from the Kondo insulator to the Mott insulator.

The singularity  of the UKLM at $J = 0$ may be 
alleviated by including an antiferromagnetic coupling between the localized
$f$-spins. In the large $U/t$ limit this leads to a bilayer spin model 
which has been considered by Vojta and Becker \cite{Vojta99}.  
The authors arrive at a similar conclusion namely that hole dynamics are 
governed by local spin environment.  Numerical work on this modified model
is presently under progress.

Given our results it is very tempting to speculate on the effects of doping with
a finite density of {\em holes} $n_h$ away from half filling. In the limit $J/t
\rightarrow \infty $ the Kondo lattice model can be mapped onto a Hubbard model with
a Coulomb repulsion $\tilde{u} \rightarrow \infty$ and a 
{\it particle } density $n_h$ 
\cite{Lacroix85}. In this low-density limit single particle renormalizations
\cite{Fetter} may be neglected which suggests that doping
the UKLM can be understood approximately within a rigid-band
picture. From this we would conclude that off half filling the UKLM
displays a Fermi surface centered around $\vec{k}=(\pm\pi,\pm\pi)$
for all values of $U$ and $J > 0$.

\end{section}


\begin{thebibliography}{10}

\bibitem{Brinkman70}
W.~F. Brinkman and T.~M. Rice, Phys. Rev. B {\bf 2},  1324  (1970).

\bibitem{Imada_rev}
M. Imada, A. Fujimori, and Y. Tokura, Rev. Mod. Phys. {\bf 70},  1039  (1998).

\bibitem{Tsunetsugu97_rev}
H. Tsunetsugu, M. Sigrist, and K. Ueda, Rev. Mod. Phys. {\bf 69},  809  (1997).

\bibitem{Assaad99a}
F.~F. Assaad, Phys. Rev. Lett. {\bf 83},  796  (1999).

\bibitem{Capponi00}
S. Capponi and F.~F. Assaad, Phs. Rev. B {\bf 63},  155113  (2001).

\bibitem{Feldbach00}
M. Feldbacher and F.~F. Assaad, Phys. Rev. B {\bf 63},  73105  (2001).

\bibitem{Jurecka01}
C. Jurecka and W. Brenig, cond-mat/0103511  .

\bibitem{Sachdev90}
S. Sachdev and R.~N. Bhatt, Phys. Rev. B {\bf 41},  9323  (1990).

\bibitem{Eder97}
R. Eder, Y. Ohta, and G.~A. Sawatzky, Phys. Rev. B {\bf 55},  3414  (1997).

\bibitem{Eder98}
R. Eder, O. Rogojanu, and G.~A. Sawatzky, Phys. Rev. B {\bf 58},  7599  (1998).

\bibitem{Normand97}
B. Normand and T. Rice, Phys. Rev. B {\bf 56},  8760  (1997).

\bibitem{Doniach77}
S. Doniach, Physica B {\bf 91},  231  (1977).

\bibitem{Matsushita97}
Y. Matsushita, M.~P. Gelfand, and C. Ishii, J. Phs. Soc. Jpn. {\bf 66},  1153
  (1997).

\bibitem{Jarrell96}
M. Jarrell and J. Gubernatis, Physics Reports {\bf 269},  133  (1996).

\bibitem{Brunner00b}
M. Brunner, F.~F. Assaad, and A. Muramatsu, Phys. Rev. B {\bf 62},  12395
  (2000).

\bibitem{Preuss95}
R. Preuss, W. Hanke, and W. von~der Linden, Phys. Rev. Lett. {\bf 75},  1344
  (1995).

\bibitem{Vojta99}
M. Vojta and K.~W. Becker, Phys. Rev. B {\bf 60},  15201  (1999).

\bibitem{Lacroix85}
C. Lacroix, Solid Stat. Commun. {\bf 54},  991  (1985).

\bibitem{Fetter}
A.~L. Fetter and J. Walecka, {\em Quantum theory of many-particle systems}
  (McGraw-Hill, New-York, 1971).

\end{thebibliography}

\end{document}